# Oxidation of copper during physical sputtering deposition: mechanism, avoidance and utilization

Jiangbin Su[1,2,*], Yang Liu[1], Meiping Jiang[1,‡], Xianfang Zhu[2]

*1. Experiment Center of Electronic Science and Technology & Department of Electronic Science and Engineering, School of Mathematics and Physics, Changzhou University, Changzhou 213164, PR China*

*2. China-Australia Joint Laboratory for Functional Nanomaterials & Physics Department, Xiamen University, Xiamen 361005, PR China*

*,‡ Corresponding authors. Tel.: +86 051986330300. E-mail addresses: jbsu@cczu.edu.cn (J.B. Su); mpjiang@cczu.edu.cn (M.P. Jiang).

**Abstract:** In this paper, oxidation of Cu during physical sputtering deposition in a high purity and low pressure Ar atmosphere without introducing $O_2$ gas flow was studied systemically. It was found that various flexible Cu-based films such as pure Cu films, $Cu/Cu_2O$ composite films, pure $Cu_2O$ films, $Cu_2O/CuO$ composite films and pure CuO films could be obtained by simply adjusting deposition parameters. Electrical and optical testing results showed that the achieved pure Cu films and $Cu/Cu_2O$ composite films both presented an intriguing combination of metal and semiconductor characteristics. For pure Cu films, the electrical resistivity and energy gap are $\sim 10^{-4}$ $\Omega \cdot cm$ and 1.54 eV respectively; while for $Cu/Cu_2O$ composite films, the electrical resistivity and energy gap are $\sim 10^{-5}$ $\Omega \cdot cm$ and 2.48 eV respectively. It is expected that such Cu-based films with a superior conductivity and a solar-window bandgap may have fascinating potential applications such as in high electron mobility transistors, electrodes and solar cells. Further, the oxidation mechanisms of Cu under different deposition parameters and the main $O_2$ source during physical sputtering deposition were also explored.

## 1. Introduction

Copper (Cu) thin films have been regarded as the next material to replace aluminum (Al) metallization in ultra-large-scale integrated circuits (ULSIC) due to their noticeable advantages such as low resistivity and superior electro-migration resistance [1,2]. In literature, the study of Cu films has been generally centered on the optimization of film properties under various deposition techniques [3-5] for specific applications such as interconnection candidate of ULSIC. However, we should further note that the surface of Cu films especially of nanoscale size is susceptible to oxygen and gets oxidized spontaneously not only when exposing to the air but also during film growing. For the former cause, we can package the Cu interconnection-based chips to keep away from the air and water as much as possible; for the latter cause, we often try to deposit Cu films in a pre-pumped high vacuum chamber with (e.g. physical sputtering) or without (e.g. thermal or electron-beam evaporation) a high-purity working rare gas such as argon (Ar). In physical sputtering deposition, the base pressure of chamber is generally up to $10^{-4}$ - $10^{-5}$ Pa and the working gas is often of 99.999 wt.% high purity [5-6]. As presented in this paper, however, we experimentally demonstrate that even under such a high pure condition Cu films can be partially or even totally oxidized during film growth (see further). Obviously, such undesirable oxidation would influence the performance, stability and lifetime of the Cu films used as interconnection in ULSIC to a certain extent or even severely. Thus, study on the oxidation of Cu during physical sputtering deposition is crucial not only to mechanism understanding but also to practical avoidance of the undesirable oxidation.

On the other hand, as the most common two kinds of Cu oxides, tenorite (CuO) and cuprite ($Cu_2O$) have attracted much attention due to their applications in low-cost solar cells and other optoelectronic devices [7-8]. However, there is still little research focused on the fabrication and properties of Cu-based composite films such as $Cu/Cu_2O$ films, which comprise of metal and semiconductor components and may have different properties and potential applications. In addition, the commonly-used sputtering deposition technology for the fabrication of CuO and $Cu_2O$ is reactive sputtering deposition [8] rather than physical sputtering deposition as applied in this paper. Different from physical sputtering deposition, additional reactive gas flow such as high pure oxygen ($O_2$) is used in reactive sputtering deposition to supply $O_2$ source for the oxidation reactions of sputtering material atoms. However, according to our calculation results (see further), it is found that the residual $O_2$ in the pre-pumped high vacuum chamber ($5.0 \times 10^{-4}$ Pa) and the high purity and low pressure sputtering gas Ar (99.999 wt.%, 0.1 Pa) is sufficient for full oxidation of sputtered Cu atoms. Thus, it is expected that physical sputtering of Cu in high pure Ar atmosphere can also be utilized to fabricate Cu-based oxide films. Chandra *et al* [9] have demonstrated that $Cu_2O$ thin films could be synthesized without using external $O_2$ source but at a high pressure of inert gas (at and above 100 mTorr). Although the $O_2$ source during high pressure sputtering was not revealed in Ref. [9], different from our low pressure case, it should originate from the impurity of inert gas rather than the residual $O_2$ in the pre-pumped high vacuum according to our calculation. Furthermore, the parameter-dependent oxidation behavior and mechanism of Cu during physical sputtering deposition especially at a commonly-used low pressure have not been studied systematically.

With the above considerations, in this paper we particularly studied the oxidation of Cu during physical sputtering deposition in a high purity and low pressure Ar atmosphere. Our calculation results showed that the main $O_2$ source for the oxidation of Cu was the residual $O_2$ in chamber vacuum which originated from the limited pre-pumped vacuum along with the impurity of working gas. Thermodynamic and kinetic analyses demonstrated that the oxidation of Cu could work spontaneously but depended on deposition parameters greatly. Further, we studied the fabrication of Cu-based oxide films by physical sputtering deposition and their electrical and optical properties. It was found that various flexible Cu-based films such as pure Cu films, $Cu/Cu_2O$ composite films, pure $Cu_2O$ films, $Cu_2O/CuO$ composite films and pure CuO films could be obtained by simply controlling the deposition parameters. Electrical and optical testing results showed that the achieved $Cu_2O$ films displayed a typical semiconductor characteristic while the achieved pure Cu thin films and $Cu/Cu_2O$ composite films both presented an intriguing combination of metal and semiconductor characteristics.

## 2. Experimental

All Cu and Cu-based samples shown in this paper were deposited on glass slides in a JGP500A balanced magnetron sputtering system. The Cu target was of 99.99 wt.% purity and the glass slide substrates were cleaned following standard cleaning procedures. The distance between the target and the substrates was set to be ~15 cm. A base pressure of $5.0 \times 10^{-4}$ Pa was achieved by a turbo-molecular pump backed by a mechanical pump. Before each deposition, a pre-sputtering of 10 min was regularly performed to remove the possible oxide layers on the target surface. Then physical sputtering deposition was carried out at a pressure of 0.1 Pa in a high pure Ar atmosphere (purity 99.999 wt.%, flow rate 15 sccm). To explore the oxidation behavior of Cu during film growing, we designed five typical sets of experiments by changing deposition parameters such as work mode (direct current and radio frequency, DC & RF), sputtering power density, substrate bias voltage and substrate temperature. As different sputtering deposition conditions are often with different deposition rates, the deposition rate and time were monitored by a quartz crystal oscillator near the substrate to make each set comparable. The detailed values of these parameters can be found in Table 1. Since the glass slide substrates are mainly composed of silicon and oxygen elements, the common energy dispersive X-ray spectroscopy (EDX) analysis is not suitable for the detection of oxygen element in the glass slide-substrated thin films. Alternatively, a powder X-ray diffractometer (XRD, RIGAKU D/Max 2500 PC) was applied subsequently for the composition or phase characterization of the as-deposited films. This is because the sputter deposits always tend to crystallize during deposition even at room temperature, which can be well detected by XRD. Further, electrical and optical testing experiments were carried out by using a four-point probe instrument (SDY-4) and an ultraviolet-visible (UV-Vis) spectrophotometer (SHIMADZU UV-2450).

## 3. Results and discussion

The XRD patterns in Figs. 1-5 show the compositions or phases of the physically sputtered Cu deposits under different deposition parameters including work mode (DC & RF), sputtering power density $p$, substrate bias voltage $V_s$ and substrate temperature $T_s$. (1) For different work modes, as shown in Fig. 1, the deposit achieved by DC mode displays a pure composition of Cu while the deposit achieved by RF mode displays a mixed oxidation composition of $CuO/Cu_2O$; (2) for different sputtering power densities $P$, as shown in Fig. 2, the deposit obtained at $p = 2.19$ W/cm$^2$ exhibits a pure composition of Cu while the deposit obtained at $p = 0.88$ W/cm$^2$ exhibits a partial oxidation composition of $Cu/Cu_2O$; (3) for different work modes and sputtering power density, as shown in Fig. 3, the deposit gained at RF 1.76 W/cm$^2$ demonstrates a single oxidation composition of $Cu_2O$ while the deposit gained at DC 0.88 W/cm$^2$ demonstrates a partial oxidation composition of $Cu/Cu_2O$; (4) for different substrate bias voltages $V_s$, as shown in Fig. 4, the deposit gotten at $V_s$ = 0 V presents a mixed oxidation composition of CuO/Cu$_2$O while the deposit gotten at $V_s$ = -100 V presents a single oxidation composition of CuO; (5) for different substrate temperatures $T_s$, as shown in Fig. 5, the deposit acquired at $T_s$ = 20 °C (room temperature) shows a pure composition of Cu while the deposit acquired at $T_s$ = 400 °C shows a partial oxidation composition of Cu/Cu$_2$O. The deposition parameters and the corresponding XRD analysis results are further listed in Table 1. The same experiments were repeated several times and similar XRD results were obtained. We can thus conclude that DC sputtering mode, high sputtering power density, low substrate bias voltage and low substrate temperature are more suitable for the fabrication of pure Cu films; on the contrary, RF sputtering mode, low sputtering power density, high substrate bias voltage and high substrate temperature are more possible to prepare Cu films with Cu$_2$O component or even total Cu$_2$O and/or CuO films.

As experimentally demonstrated above, partial or even entire sputtered Cu atoms can be oxidized into Cu$_2$O or CuO during physical sputtering deposition. Generally, there are three possible O$_2$ sources applied for the oxidation of Cu: (1) the impurity of 99.99 wt.% Cu target; (2) the O$_2$ molecules adsorbed on the glass slide surface; (3) the limited pre-pumped vacuum (5.0×10$^{-4}$ Pa) and the impurity of 99.999 wt.% and 0.1 Pa Ar working gas. Alternatively, someone may think that the release of O$_2$ adsorbed on the inner wall of chamber would also supply the O$_2$. Actually, such release of O$_2$ should be and is considered as one part of the 3$^{rd}$ source (the remained O$_2$ in pre-pumped vacuum) because the O$_2$ after releasing would come into the chamber vacuum.

In the following, we analyze and compare the respective contents of O and Cu at the above mentioned three specified locations. For the 1$^{st}$ possible source, assuming that all the impurity in the Cu target is O element, the atomic ratio of O and Cu is 1:2500. So small is the O content that we can completely ignore its influence. For the 2nd possible source, we compare the surface density of adsorbed $O_2$ on glass slide surface with that of as-deposited Cu atomic layer. According to ref. [10], we know that the adsorption of $O_2$ on glass (total surface: 1966 cm$^2$) at a pressure of 0.7 torr is 0.63 mm$^3$ (volume at 20 °C and 760 mm pressure). Since the quantity of adsorbed gas is proportional to the pressure at low pressure [10], the surface density of adsorbed $O_2$ on glass slide surface in our case is about $2.2\times10^5$ atoms/cm$^2$. In comparison, the surface density of as-deposited Cu about $10^{14}$ atoms/cm$^2$ is much larger. Accordingly, we can ignore the little influence of adsorbed $O_2$ on the surface of glass slide substrates. For the 3rd possible source, the partial oxygen pressure in chamber vacuum (including two parts: one is from the pre-pumped high vacuum, and the other one is from the high pure working gas Ar) is about $9.5\times10^{-5}$ Pa. In this calculation, for the sake of simplicity we assume that the volume fraction of $O_2$ in the pre-pumped high vacuum is 21%, which is similar to the composition of air. Due to the low pressure, we can apply the State Equation of Ideal Gas (i.e. $PV = nRT$) to obtain the concentration of $O_2$ in chamber vacuum. Our calculation result is about $4.8\times10^{10}$ atoms/cm$^3$. On the other hand, provided no or little reflection occurs for simplicity during the incident Cu flux striking the substrate, we can achieve the following equation:

$$n_{Cu} = \frac{\rho_{Cu} v_d N_A}{vM} \quad (1)$$

where $n_{Cu}$ is the concentration of incident Cu atoms near the glass substrate, $\rho_{Cu}$ is the density of Cu film about 8.4 g/cm$^3$ [6], $v_d$ is the deposition rate of Cu atoms with a typical value of 0.01 - 0.2 nm/s in this paper, $N_A$ is the Avogadro constant, $v$ is the average velocity of travelling Cu atoms before reaching the substrate, $M$ is the atomic weight of Cu. After a detailed calculation as ref. [11], we find that the average retained kinetic energy of a sputtered Cu atom in this work is 1.24 eV. Thus the average velocity $v$ of Cu atoms near the substrate is about $1.9\times10^3$ m/s. According to Eq. (1), the concentration of incident Cu atoms near the substrate is about $10^8$ - $10^9$ atoms/cm$^3$. Since the concentration of $O_2$ in chamber vacuum is larger than that of Cu atoms near the substrate ($4.8\times10^{10}$ vs. $10^8$ - $10^9$ atoms/cm$^3$), the residual $O_2$ is sufficient in quantity and should be the main source for the oxidation of Cu atoms during physical sputtering deposition. What is more, part of sputtered Cu atoms may be preferentially oxidized on the target surface before travelling to the substrate. The comparison of O and Cu contents at the above mentioned three different locations is further shown in Table 2.

Based on thermodynamic data, the reactions between Cu and $O_2$ are as follows.

$$2Cu + \frac{1}{2}O_2(g) \to Cu_2O \qquad (2)$$

$$Cu + \frac{1}{2}O_2(g) \to CuO \qquad (3)$$

$$\frac{1}{2}Cu_2O + \frac{1}{4}O_2(g) \to CuO \qquad (4)$$

The changes in free energy at room temperature given by Eqs. (2-4) are -147.69, -128.12 and -54.27 kJ·mol$^{-1}$ respectively, all of which are much lower than zero. It indicates that Cu can be oxidized spontaneously at room temperature under an atmosphere containing sufficient $O_2$. We should further note that, different from the oxidation of bulk Cu, the oxidation of gaseous Cu atoms during sputtering deposition would be greatly enhanced due to a much higher energy and a much more full exposure to $O_2$. In addition, the oxidation of Cu into $Cu_2O$ (Eq. (2)) seems the easiest to occur due to the biggest change in free energy. In spite of this, Cu and $Cu_2O$ can also be oxidized or further oxidized into CuO due to the changes in free energy given by Eqs. (3-4) are both much lower than zero.

Thermodynamically, as demonstrated above, the sputtered Cu atoms can react spontaneously at room temperature with the sufficient $O_2$ molecules retained in vacuum. Nevertheless, not all the sputter deposits are $Cu_2O$ or $CuO$ as expected, which are observed to be dependent kinetically on the deposition parameters such as work modes, sputtering power density, substrate bias voltage and substrate temperature (see Figs. 1-5 or Table 1). In the following, we discuss the influences of the above deposition parameters on the oxidation of Cu atoms during physical sputtering deposition.

For different work modes, DC mode relative to RF mode leads to a much higher deposition rate (see Set 1, Table 1). Similarly, as shown in Set 2 of Table 1, larger sputtering power density could also cause a much higher deposition rate. Generally speaking, a higher deposition rate could shorten the required deposition time which would prevent Cu atoms from sufficiently exposing to $O_2$. This should be the main reason why DC mode with large sputtering power density tends to achieve pure Cu films while RF mode or small sputtering power density tends to form Cu films with $Cu_2O$ component or even total $Cu_2O$ and/or $CuO$ films. However, relative to DC 0.88 $W/cm^2$, as shown in Set 3 of Table 1, RF 1.76 $W/cm^2$ with a higher deposition rate (0.045 vs. 0.03 nm/s) and a shorter period of deposition time (1.5 vs. 2 h) seems to obtain a more fully oxidized film ($Cu_2O$ vs. $Cu/Cu_2O$) inversely. It indicates that there may be another factor influencing the oxidation of Cu besides the deposition rate and time. It is expected that the different work modes DC and RF result in such different oxidation effect. As is known to all, similar to the working gas Ar, the residual $O_2$ between two electrodes would be ionized into plasma (including $O_2^+$, $e^-$, O, $O_2$, etc.) during sputtering. For DC sputtering, the $O_2^+$ travels directionally to the target surface (cathode). While for RF sputtering, the $O_2^+$ travels back and forth between two electrodes which may increase the opportunity of reaction with the sputtered Cu atoms. For different substrate bias voltage or different substrate temperature, as shown in Sets 4 and 5 of Table 1, the work mode, deposition rate and time between two typical experiments are the same or nearly the same. However, the achieved sputter deposits are more fully oxidized under higher substrate bias voltage (CuO vs. CuO/Cu$_2$O) or higher substrate temperature (Cu/Cu$_2$O vs. Cu). This is probably due to the additional energy supplied to Cu atoms and O$_2$ molecules under higher substrate bias voltage or higher substrate temperature promotes the oxidation of Cu or further oxidation of Cu$_2$O.

Although we can try to pump the chamber vacuum to less than 10$^{-6}$ Pa to avoid the undesirable oxidation as much as possible, it would greatly increase the fabrication cost and time. What is worse, the common high-pure working gas Ar is typically of 99.999 wt.% purity, which would also introduce unexpected O$_2$ and make it infeasible. Alternatively in this paper, DC mode with a high sputtering power density such as 2.19 W/cm$^2$ is demonstrated to be an efficient and low-cost method to deposit pure Cu films. Further, it is also demonstrated that we can deposit Cu films with Cu$_2$O component or Cu$_2$O and/or CuO films by adjusting the deposition parameters. For example, DC mode with a low sputtering power density such as 0.88 W/cm$^2$ at room temperature can be applied to fabricate Cu/Cu$_2$O composite films; DC mode with a high sputtering power density such as 2.19 W/cm$^2$ at 400°C can also be applied to fabricate Cu/Cu$_2$O composite films. RF mode with a strong magnetic field of ~6 kGs (see caption of Table 1) and a sputtering power density of 1.76 W/cm$^2$ is suitable for the fabrication of Cu$_2$O films; while RF mode with a relatively weaker magnetic field of ~4.5 kGs (see caption of Table 1) and a sputtering power density of 2.19 W/cm$^2$ is suitable for the fabrication of CuO films (with substrate bias voltage) or Cu$_2$O/CuO composite films (without substrate bias voltage).

Table 3 and Fig. 6 show the electrical resistivity and the $(Ahv)^2$ - $hv$ curves of the achieved Cu$_2$O, Cu and Cu/Cu$_2$O films respectively. For Cu$_2$O film, the electrical resistivity is larger than 10$^0$ Ω·cm and the energy gap is 2.06 eV, which shows a typical semiconductor characteristic; for pure Cu film, it displays an intriguing combination of metal and semiconductor characteristics with a low electrical resistivity of ~$10^{-4}$ Ω·cm and a narrow bandgap of 1.54 eV; while for Cu/Cu$_2$O composite film, even more intriguing, it also displays a combination of metal and semiconductor characteristics with a much lower electrical resistivity of ~$10^{-5}$ Ω·cm and a much wider bandgap of 2.48 eV. It is expected that quantum size effect of two-dimensional thin films (and the massive doping of Cu in Cu$_2$O for Cu/Cu$_2$O composite film) results in such interesting phenomena although the detailed mechanisms are still unclear. In spite of this, such pure Cu films and Cu/Cu$_2$O composite films with a low electrical resistivity and a bandgap at the solar window may have fascinating potential applications such as in high electron mobility transistors [12], electrodes [13] and solar cells [14-15].

## 4. Conclusions

In this work, the oxidation of Cu during physical sputtering deposition was studied by varying the deposition parameters including work mode, sputtering power density, substrate bias voltage, and substrate temperature. Our calculation results showed that the main O$_2$ source for the oxidation of Cu was the residual O$_2$ in chamber vacuum which originated from the limited pre-pumped vacuum ($5.0\times10^{-4}$ Pa) along with the impurity of 99.999 wt.% and 0.1 Pa working gas Ar. It was demonstrated that DC mode with a large sputtering power density such as 2.19 W/cm$^2$ at room temperature tended to deposit pure Cu films, while DC mode with a small sputtering power density such as 0.88 W/cm$^2$ or at an elevated temperature of 400°C tended to deposit Cu/Cu$_2$O composite films. It was also demonstrated that RF mode normally with a low deposition rate was more suitable for the fabrication of Cu$_2$O and/or CuO films rather than pure Cu or Cu/Cu$_2$O composite films. Electrical and optical testing results showed that the achieved Cu$_2$O films displayed a typical semiconductor characteristic with an electrical resistivity larger than $10^0$ Ω·cm and an energy gap of 2.06 eV. However, the achieved pure Cu films and Cu/Cu$_2$O composite films both presented an intriguing combination of metal and semiconductor characteristics. For pure Cu films, the electrical resistivity and energy gap are ~$10^{-4}$ Ω·cm and 1.54 eV respectively; while for Cu/Cu$_2$O composite films, the electrical resistivity and energy gap are ~$10^{-5}$ Ω·cm and 2.48 eV respectively. Based on these findings, it provided not only an effective route to avoid the oxidation of sputtered Cu atoms during physical sputtering deposition but also a flexible and low-cost new method to prepare various Cu-based oxide films.

## Acknowledgements

The authors thank Honghong Wang, Dingjuan Pan and Dan Jin at School of Mathematics and Physics, Changzhou University for their experimental assistance.

## References

[1] C.W. Park, R.W. Vook, Activation energy for electromigration in Cu films, Appl. Phys. Lett. 59 (1991) 175-177.

[2] S.P. Murarka, R.J. Gutmann, Advanced multilayer metallization schemes with copper as interconnection metal, Thin Solid Films 236 (1993) 257-266.

[3] R.A. Roy, J.J. Cuomo, D.S. Yee, Control of microstructure and properties of copper films using ion-assisted deposition, J. Vac. Sci. Technol. A 6 (1988) 1621-1626.

[4] W.M. Holber, J.S. Logan, H.J. Grabarz, et al, Copper deposition by electron cyclotron resonance plasma, J. Vac. Sci. Technol. A 11 (1993) 2903-2910.

[5] K.Y. Chan, T.Y. Tou, B.S. Teo, Effects of substrate temperature on electrical and structural properties

of copper thin films, Microelectron J. 37 (2006) 930-937.

[6] H.M. Choi, S.K. Choi, O. Aderson, et al, Influence of film density on residual stress and resistivity for Cu thin films deposited by bias sputtering, Thin Solid Films 358 (2000) 202-205.

[7] B. Balamurugan, B.R. Mehta, Optical and structural properties of nanocrystalline copper oxide thin films prepared by activated reactive evaporation, Thin Solid Films 396 (2001) 90-96.

[8] A. Parretta, M.K. Jayaraj, A.D. Nocera, et al, Electrical and optical properties of copper oxide films prepared by reactive RF magnetron sputtering, Phys. Stat. Sol. 155 (1996) 399-404.

[9] R. Chandra, P. Taneja, P. Ayyub, Optical properties of transparent nanocrystalline $Cu_2O$ thin films synthesized by high pressure gas sputtering, Nanostruct. Mater. 11 (1999) 505-512.

[10] I. Langmuir, The adsorption of gases on plane surfaces of glass, mica and platinum, J. Am. Chem. Soc. 40 (1918) 1361-1403.

[11] J.B. Su, X.X. Li, M.P. Jiang, et al, Layer-plus-wire growth of copper by small incident angle deposition, Mater. Lett. 92 (2013) 304-307.

[12] S. Balendhran, J. Deng, J.Z. Ou, et al, Enhanced charge carrier mobility in two-dimensional high dielectric molybdenum oxide, Adv. Mater. 25 (2013) 109-114.

[13] S. Anandan, X. Wen, S. Yang, Room temperature growth of CuO nanorod arrays on copper and their application as a cathode in dye-sensitized solar cells, Mater. Chem. Phys. 93 (2005) 35-40.

[14] J.J. Loferski, Theoretical considerations governing the choice of the optimum semiconductor for photovoltaic solar energy conversion, J. Appl. Phys. 27 (1956) 777-784.

[15] H. Tanaka, T. Shimakawa, T. Miyata, et al, Effect of AZO film deposition conditions on the photovoltaic properties of AZO-$Cu_2O$ heterojunctions, Appl. Surf. Sci. 244 (2005) 568-572.

**Table captions**

**Table 1** Five typical sets of experiments under different deposition parameters, where $p$ is the sputtering power density, $V_s$ is the substrate bias voltage, $T_s$ is the substrate temperature, $v_d$ and $t$ are the deposition rate and time respectively. Remark: Different from the RF mode in Sets 1 and 4, the RF' mode in Set 3 with a stronger magnetic field (~6 vs. ~4.5 kGs) causes a higher deposition rate (0.045 vs. 0.02 nm/s) even under a lower sputtering power density (1.76 vs. 2.19 W/cm$^2$).

**Table 2** The contents of O and Cu elements at different locations.

**Table 3** Electrical resistivity of the achieved Cu$_2$O, Cu and Cu/Cu$_2$O films.

**Figure captions**

**Fig. 1** XRD patterns showing the composition of the Cu deposits obtained under different work modes (Set 1: DC & RF 2.19 W/cm$^2$).

**Fig. 2** XRD patterns showing the composition of the Cu deposits obtained under different sputtering power density (Set 2: $p$ = 2.19 W/cm$^2$, 0.88 W/cm$^2$).

**Fig. 3** XRD patterns showing the composition of the Cu deposits obtained under different work modes and sputtering power density (Set 3: RF 1.76 W/cm$^2$, DC 0.88 W/cm$^2$).

**Fig. 4** XRD patterns showing the composition of the Cu deposits obtained under different substrate bias voltage (Set 4: $V_s$ = 0 V, -100 V).

**Fig. 5** XRD patterns showing the composition of the Cu deposits obtained under different substrate temperature (Set 5: $T_s$ = 20 °C, 400 °C).

**Fig. 6** $(Ahv)^2$ - $hv$ curves of the achieved (a) Cu$_2$O, (b) Cu and (c) Cu/Cu$_2$O films.

# Tables

Table 1:

| Sets | Work mode | $p$ (W/cm$^2$) | $V_s$ (V) | $T_s$ (°C) | $v_d$ (nm/s), t (h, s) | Phase |
|---|---|---|---|---|---|---|
| 1 | **DC** | 2.19 | 0 | 20 | ~0.2, 155s | Cu |
|   | **RF** | 2.19 | 0 | 20 | ~0.02, 1.5h | CuO, Cu$_2$O |
| 2 | DC | **2.19** | 0 | 20 | ~0.2, 288s | Cu |
|   | DC | **0.88** | 0 | 20 | ~0.03, 2h | Cu, Cu$_2$O |
| 3 | **RF'** | 1.76 | 0 | 20 | ~0.045, 1.5h | Cu$_2$O |
|   | **DC** | 0.88 | 0 | 20 | ~0.03, 2h | Cu, Cu$_2$O |
| 4 | RF | 2.19 | **0** | 20 | ~0.02, 1.5h | CuO, Cu$_2$O |
|   | RF | 2.19 | **-100** | 20 | ~0.02, 1.5h | CuO |
| 5 | DC | 2.19 | 0 | **20** | ~0.2, 155s | Cu |
|   | DC | 2.19 | 0 | **400** | ~0.2, 155s | Cu, Cu$_2$O |

Table 2:

| Locations | O | Cu |
|---|---|---|
| 1$^{st}$: Cu target (ratio) | 1 | 2500 |
| 2$^{nd}$: glass slide surface | 2.2×10$^5$ atoms/cm$^2$ | ~10$^{14}$ atoms/cm$^2$ |
| 3$^{rd}$: vacuum near substrate | 4.8×10$^{10}$ atoms/cm$^3$ | ~10$^8$ - 10$^9$ atoms/cm$^3$ |

Table 3:

| Samples | Electrical Resistivity |
|---|---|
| $Cu_2O$ film | $>10^0$ Ω·cm |
| Cu film | $\sim10^{-4}$ Ω·cm |
| $Cu/Cu_2O$ film | $\sim10^{-5}$ Ω·cm |

# Figures

Fig. 1:

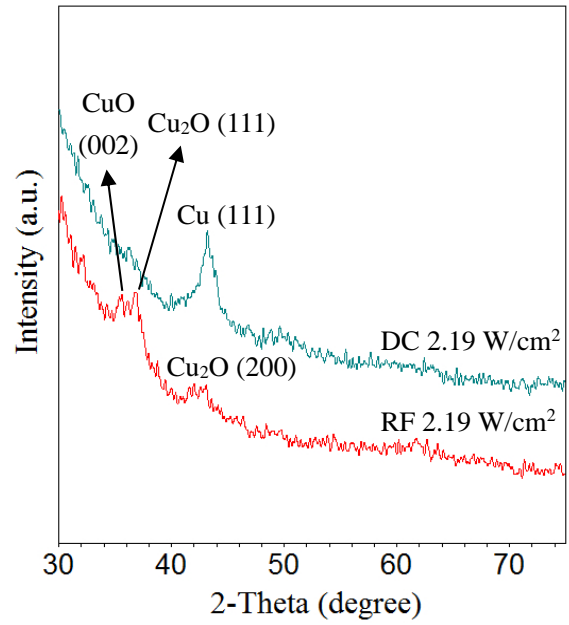

Fig. 2:

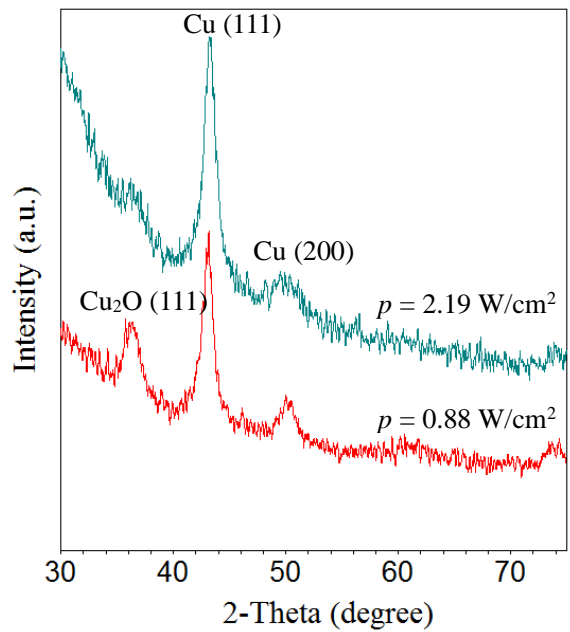

Fig. 3:

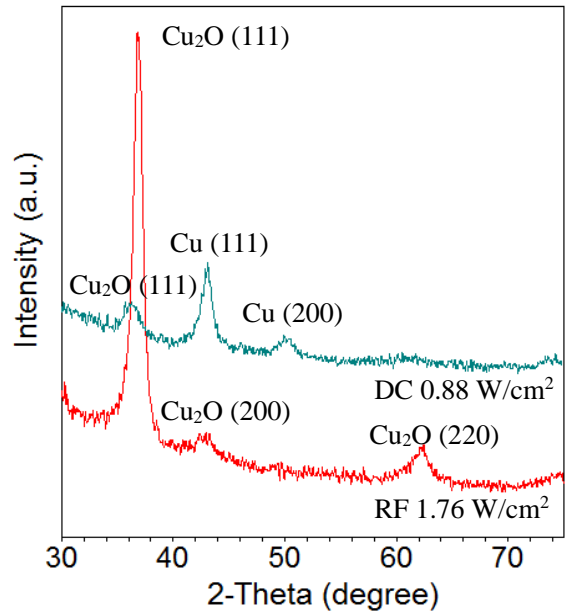

Fig. 4:

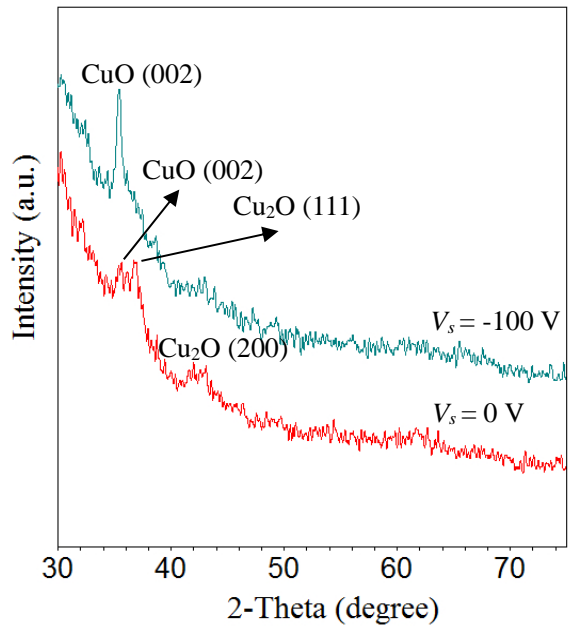

Fig. 5:

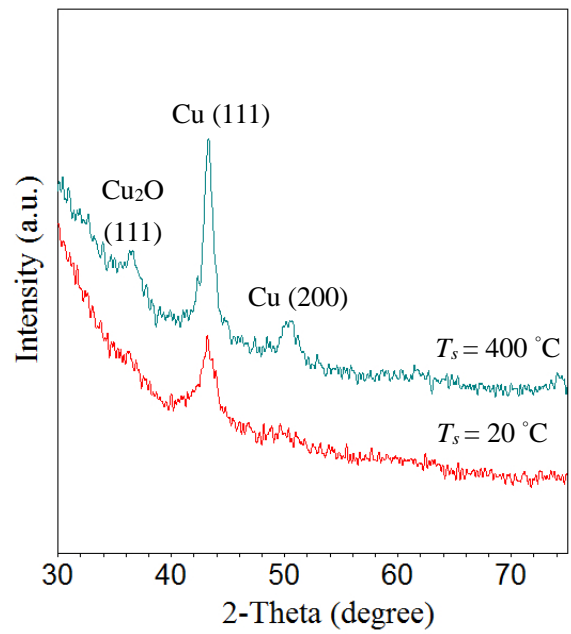

Fig. 6:

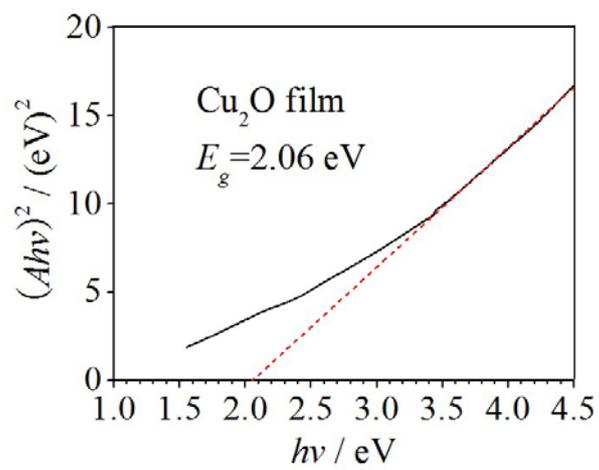

(a)

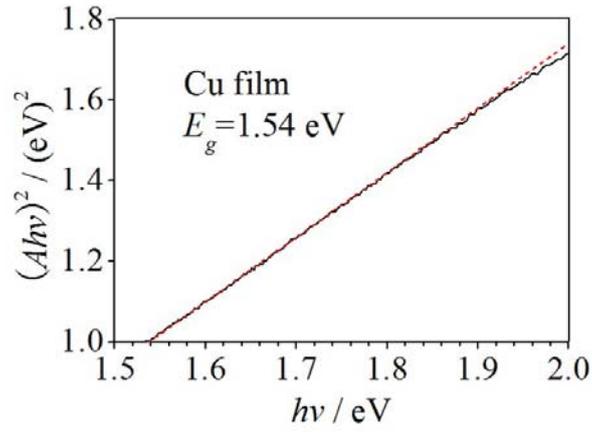

(b)

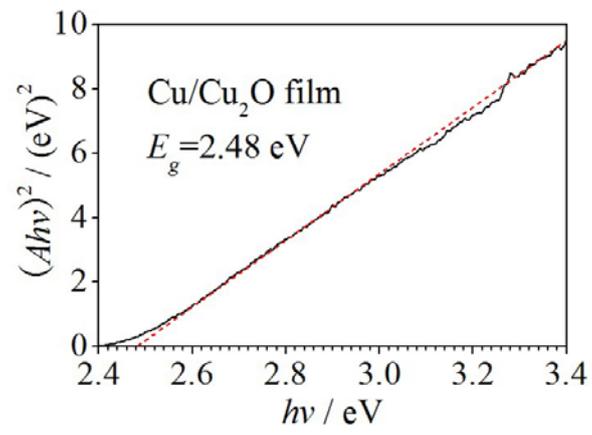

(c)